\documentclass[aps,prl,twocolumn,showpacs,amsmath,amsmath,amssymb,amsfonts]{revtex4-1}
\usepackage{graphicx}% Include figure files
\usepackage{dcolumn}% Align table columns on decimal point
\usepackage{bm}% bold math
\usepackage{color}
\usepackage{multirow}

\begin{document}
\author{Weibin Li}
\author{Igor Lesanovsky}
\affiliation{Midlands Ultracold Atom Research Centre (MUARC), School
of Physics and Astronomy, The University of Nottingham, Nottingham,
NG7 2RD, United Kingdom}
\title{Electronically excited cold ion crystals}
\date{\today}
\begin{abstract}
The laser excitation of an ion crystal to high
lying and long-lived electronic states is a genuine many-body process even if in fact only a single ion is excited. This is a direct manifestation of the strong coupling between internal and external dynamics and becomes most apparent in the vicinity of a structural phase transition. Here we show that utilizing highly excited states offers a new approach to the coherent manipulation of ion crystals. This permits the study of phenomena which rely on a strong coupling between electronic and vibrational dynamics and opens up a route towards the quantum simulation of molecular processes in a Paul trap.
\end{abstract}


\maketitle

Cold crystals of trapped ions are currently among the most precisely controllable quantum systems and are widely used, e.g. in quantum computation \cite{cirac95,haeffner08}, optical frequency clocks \cite{chwalla09,chou10} as well as for the simulation of quantum magnets \cite{porras04,friedenauer08,kim10}, relativistic quantum mechanics \cite{gerritsma10} and open quantum systems \cite{barreiro11}. All these applications are based on the laser excitation of low lying excited states and/or on a state-dependent coupling between these electronic levels and vibrational modes of the crystal. A different regime is entered when instead ions are excited to high lying Rydberg states \cite{mueller08}. This promises applications ranging from fast quantum gates to the simulation of coherent spin dynamics and excitation transfer -- all relying on the large dipolar interaction between excited ions. The plethora of newly emerging possibilities is impressively demonstrated by recent breakthroughs in the experimental control of neutral Rydberg atoms demonstrating quantum gates \cite{urban09,gaetan09} and collective excitation dynamics \cite{viteau11}.

In this work we show that the laser excitation of a cold ion crystal to electronic Rydberg states leads to genuinely novel features which go beyond the mere extension of neutral Rydberg physics to excited ions. Even a single localized excitation exhibits a pronounced many-body character which is rooted in the strong coupling between highly excited electronic levels and vibrational modes. This induces state-dependent structural changes in the ion crystal, reminiscent of a configurational change in an excited molecule. Lifetimes of ionic Rydberg states on the order of $\sim 100\,\mu\mathrm{s}$ grant long coherence times for this `vibronic dynamics'. This highlights novel routes for the coherent manipulation of cold ion crystals and paves the way for the study of molecular phenomena with ions in the well-insulated environment of a Paul trap.
\begin{figure}
\centering
\includegraphics*[width=0.95\columnwidth]{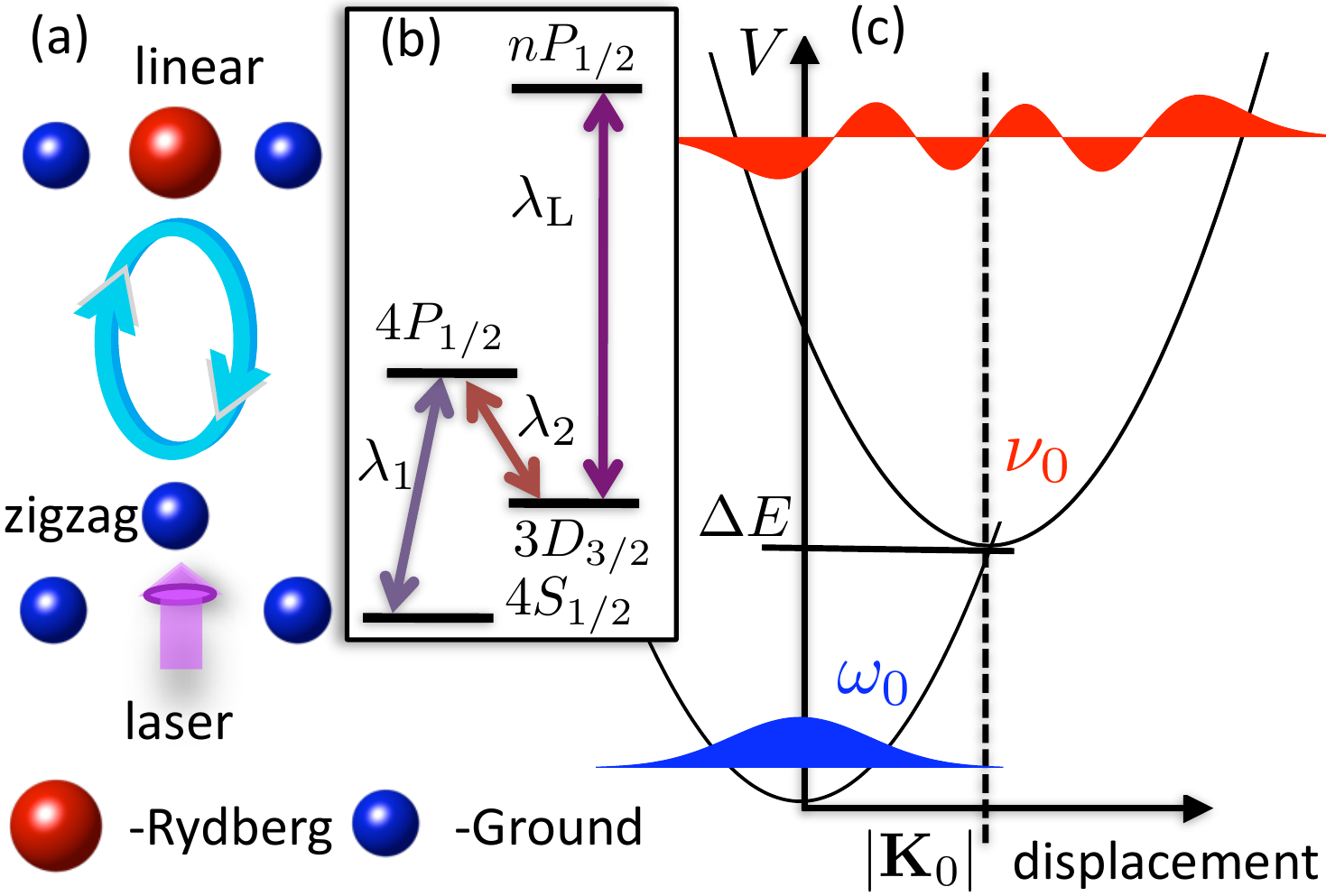}
\caption{Rydberg excitation of a three-ion crystal near a structural phase transition. (a) The trap parameters are chosen such that ground state ions (blue) form a zigzag crystal. The localized Rydberg excitation of the central ion (red) effectuates a change of the structural configuration. (b) Electronic level scheme. For the Rydberg excitation of a $^{40}$Ca$^+$ ion from the ground state $4S_{1/2}$ the ion first is transferred to the metastable state $3D_{3/2}$ via the intermediate level $4P_{1/2}$. A laser of $\lambda_{\rm{L}}\approx 122$ nm or a coherent three-photon excitation is used subsequently to populate the $nP_{1/2}$ state. (c) Sketch of the electronic potential surface of the electronically excited ion crystal and a crystal with all ions in the ground state. The $x$- and $y$-axis show the relative displacement and the heights of the two potential surfaces which can be approximated as harmonic oscillators (here with frequencies $\omega_0$ and $\nu_0$). The large mutual displacement $|\mathbf{K}_0|$ of the two surfaces, separated by the energy $\Delta E$,  indicates that the Rydberg excitation is accompanied by strong mechanical forces. This has profound implications for the laser excitation behavior of even a single ion of the crystal.}
\label{fig:scheme}
\end{figure}

The fascinating physics emerging from couplings between Rydberg levels and external motion has recently been theoretically studied for the case of neutral atoms by W\"uster \textit{et al.} \cite{Wuster10,Wuster11}. Here entanglement transfer as well as the intriguing dynamics near conical intersections of molecular potential surfaces were investigated. Both effects rely on delocalized excitations that are established by the dipole-dipole interaction among Rydberg states. While in ion crystals such a scenario can be in principle equally achieved \cite{mueller08}, the features discussed in this work rely on a different mechanism. Here the large polarizability $\sim n^7$ of a single ion excited to a Rydberg state with principle quantum number $n$ leads to a drastic change of its local trapping potential. This affects the entire crystal as changes in the position of one ion are immediately transmitted by the long-ranged Coulomb force.

The phenomenon becomes most striking in a parameter regime where the ion crystal is close to a structural phase transition. In a linear Paul trap such transition is controlled by the ratio of the axial $\omega_{\rm{z}}$ and longitudinal $\omega_{\rm{\rho}}$ trap frequencies. Depending on the so-called trapping anisotropy $\mathcal{A}=(\omega_{\rm{z}}/\omega_{\rm{\rho}})^2$, the ions in their vibrational ground state form either a linear or a zigzag crystal \cite{raizen92}. A transition between these two configurations takes place at a critical trapping anisotropy $\mathcal{A}_{\rm{c}} \approx 2.53N^{-1.73}$ \cite{schiffer93} and has been experimentally observed in crystals with $N=3...19$ ions \cite{block00}.

We are focussing here on the smallest possible system in which this transition is observable - a three-ion crystal of $^{40}$Ca$^+$. As depicted in Fig.~\ref{fig:scheme}a we start in a zigzag configuration from where we excite the middle ion to the $nP_{1/2}$ Rydberg state. Technically this is done by a standard two-step excitation to the metastable $3D_{3/2}$ state (via the state $4P_{1/2}$) followed by a coherent excitation to the Rydberg level $nP_{1/2}$ (see Fig.~\ref{fig:scheme}b). The last step can be undertaken by either a three-photon excitation or by employing a coherent source of vacuum ultraviolet light with a wavelength near $122$ nm as discussed in Ref.~\cite{schmidt11}. The excited ion will experience a state dependent - in general tighter - confinement $\omega_{\rho}(n)$ which in a qualitative picture leads to a reduction of the trap anisotropy  $\mathcal{A}(n)$ below $\mathcal{A}_{\rm{c}}$. As a consequence, the electronically excited crystal prefers a linear arrangement of the ions. The associated strong mechanical effects, illustrated by the far displaced potential surfaces depicted in Fig.~\ref{fig:scheme}b will become apparent on the spectroscopic properties of a cold ion crystal.

\begin{figure}
\includegraphics*[width=1.0\columnwidth]{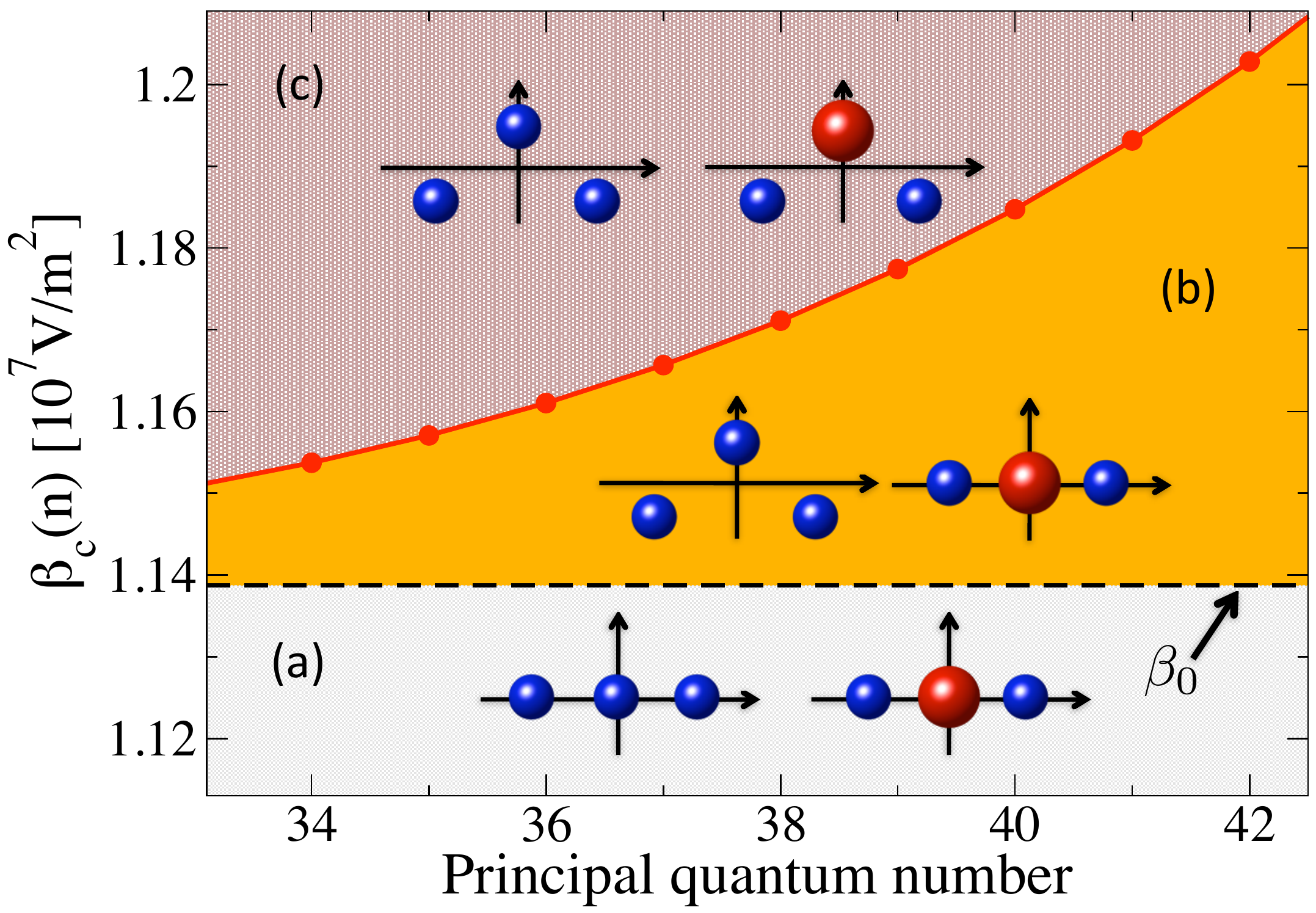}
\caption{Structural configurations of an electronically excited three-ion crystal and a crystal in which all ions are in the electronic ground state. The assumed configuration depends on the static field gradient $\beta$ and the principal quantum number $n$. While in regions (a) and (c) the configurations are identical, an electronic excitation of the central ion (red) is accompanied by a structural configuration change in region (b).}
\label{fig:criticalbeta}
\end{figure}

In the following we provide a more thorough quantitative study of this simplified picture. We consider a linear Paul trap with the electric potential $\Phi({\bf R},t)={\bf R}\cdot\bar{{\bf G}}\cdot{\bf R}^{\rm{T}}$. Here, $\bar{{\bf G}}$ is a  $3\times 3$ diagonal matrix with elements $\bar{{\bf G}}_{XX}=\alpha\cos\Omega t-\beta(1+\varepsilon),\,\bar{{\bf G}}_{YY}=-\alpha\cos\Omega t-\beta(1-\varepsilon),\,\bar{{\bf G}}_{ZZ}=2\beta$ that denote the static and time-dependent electric field gradients in the respective directions, with the latter oscillating at radio-frequency (RF) $\Omega$. In highly excited states, ions and atoms cannot be considered point-like in such an environment as the trapping field can vary significantly on the size of an electronic orbit as has been shown in Refs.~\cite{mueller08,Hezel06}. The quantum mechanical treatment of this problem has been discussed in Ref.~\cite{schmidt11} but in order to be self-contained we briefly summarize the main steps: The Hamiltonian of a single ion of mass $M$ and charge $e$ reads \cite{mueller08}
\begin{equation}
H_{\rm{ion}}({\bf R},{\bf r})=H_{\rm{c}}({\bf R}) + H_{\rm{e}}({\bf r}) +H_{\rm{et}}({\bf r}) + H_{\rm{ec}}({\bf R},{\bf r}).
\label{eq:singleion}
\end{equation}
Here $H_{\rm{c}}({\bf R})=\sum_{\xi}[P_{\xi}^2/(2M) +M\omega^2_{\xi}R_{\xi}^2/2]$ describes the harmonic motion of the center-of-mass (CM) coordinate ${\bf{R}}=(X,\,Y,\,Z)$ and the index $\xi=X,\,Y,\, Z$ refers to the respective components. This field results in the trap frequencies $\omega_X=\sqrt{2[\frac{e^2\alpha^2}{M\Omega^2}-e(1+\varepsilon)\beta]/M}$,  $\omega_Y=\sqrt{2[\frac{e^2\alpha^2}{M\Omega^2}-e(1-\varepsilon)\beta]/M}$ and  $\omega_Z=\sqrt{4e\beta/M}$. The parameter $0<\varepsilon<1$ breaks the axial symmetry of the trap which simplifies the theoretical discussion as it removes some mode degeneracies. Terms describing the coupling of the micromotion to the CM are neglected. The Hamiltonian $H_{\rm{e}}({\bf r})$ governs the dynamics of the valence electron with the relative coordinate $\mathbf{r}$. Its eigenstates $\left|\Psi_n\right>$ and eigenenergies $E_n$ (the index $n$ contains all relevant quantum numbers) are obtained from a model potential calculation \cite{aymar96}. Finally, the Hamiltonians $H_{\rm{et}}({\bf r})=-e\Phi({\bf r},t)$ and  $H_{\rm{ec}}({\bf R}, {\bf r})=-2e{\bf R}\cdot\bar{{\bf G}}\cdot{\bf r}^{\rm{T}}$ account for the electron-trap interaction and the fundamentally important coupling between the CM and electronic motion. The total Hamiltonian of the three-ion crystal in Fig.~\ref{fig:scheme}a is obtained by further including the Coulomb interaction among the ions. For the two ground-state ions this is straight-forward. In the case of the Rydberg ion the Coulomb interaction of the valence electron and the doubly charged core with the remaining two ions is accounted for by a multipole expansions \cite{mueller08}. The dominant contribution of this expansion is the dipole-charge interaction $V_{\rm{dc}}=[e^2/(4\pi\epsilon_0)]\sum_j({\bf R}-{\bf R}_j)\cdot {\bf r}/|{\bf R}-{\bf R}_j|^3$ where $\epsilon_0$ is the permittivity of vacuum, ${\bf R}$ and ${\bf r}$ denote the CM and electronic coordinate of the Rydberg ion and the vectors ${\bf R}_j$ contain the coordinates of the ground state ions.

When excited to the Rydberg state the coupling Hamiltonians $H_{\rm{ec}}$ and $V_{\rm{dc}}$ create an additional ponderomotive potential for the CM motion of the central ion. This potential provides a tighter radial confinement $V_{\rm{add}}=M\omega^2_{\rm{add}}( X^2+ Y^2)/2$ with the trap frequency $\omega_{\rm{add}}=\sqrt{\left[8(e^2\alpha^2)^2/(M\Omega^2)^2+4e^2\alpha^2\right]x^{(2)}/M}$. Here $x^{(2)}=\sum_{m\neq n}|\langle \Psi_m|x|\Psi_n\rangle|^2/(E_n-E_m)$ is proportional to the polarizability of the Rydberg electron. In deriving $V_{\rm{add}}$ we made use of the fact that the typical Kepler orbital frequency of the Rydberg electron is about $2\pi \times 100$ GHz, which is much higher than the RF frequency $\Omega$ (typically $2\pi\times 10 ... 100$ MHz) and also much higher than typical trap frequencies ($\sim2\pi\times 1$ MHz). This allows us to calculate the electronic motion within a quasistatic approximation where we treat time as a parameter \cite{schmidt11}. Compared to the vibrational dynamics of the ions, however, this surface will oscillate at the fast RF frequency. Therefore, the CM of the central ion experiences an effective ponderomotive potential.

For a quantitative discussion we choose an RF field with gradient $\alpha=10^9 \, \rm{V/m^2}$ oscillating at a frequency $\Omega=2\pi\times 30$ MHz and a parameter $\varepsilon = 0.15$ for splitting the degeneracy of the trap frequencies in $x$ and $y$ direction. For the following discussion it is more convenient to discuss the expected structural configuration change in the ion crystal as a function of the gradient $\beta$ rather than in terms of the trap anisotropy $\mathcal{A}$. The critical gradient $\beta_0$ at which the linear-zigzag transition occurs for a crystal of three ground state ions is given by
\begin{equation}
\beta_0=\frac{5}{29+5\varepsilon}\frac{e\alpha^2}{M\Omega^2}. \nonumber
\end{equation}
Due to the additional trapping potential $V_{\rm{add}}$ this critical gradient changes when the central ion is excited to a Rydberg state. This leads to a state-dependent critical gradient $\beta_c(n)$ which is shown Fig.~\ref{fig:criticalbeta} together with $\beta_0$. The plot can be thought of as a `phase diagram': in region (a) both the excited and the ground state crystal prefer a linear arrangement. A zigzag configuration is preferred in both cases when gradient and principal quantum number are chosen from region (c). In region (b) an electronic excitation is accompanied by a structural configuration change.

\begin{figure}
\centering
\includegraphics*[width=1.0\columnwidth]{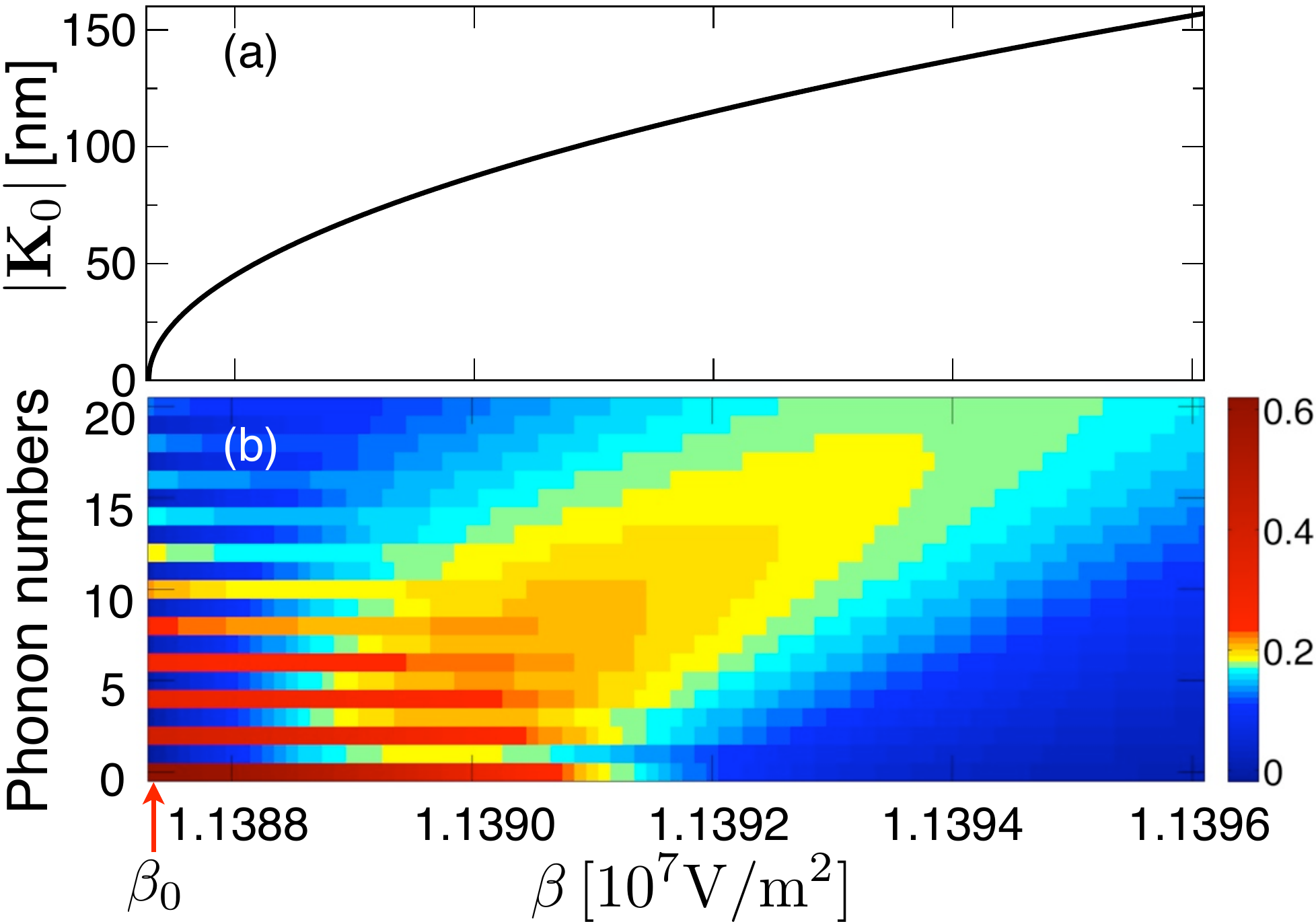}
\caption{(a) Displacement $|\mathbf{K}_0|$ of the CM potential surfaces corresponding to a ground state and and excited three-ion crystal, as illustrated in Fig.~\ref{fig:scheme}c. $|\mathbf{K}_0|$ is plotted against the trapping field gradient $\beta$ with the abscissa starting at $\beta_0$. A large displacement affects the Franck-Condon (FC) factors that quantify couplings between different vibrational states of the two potential surfaces. Panel (b) shows the FC factors for transitions from the vibrational ground state of the lower potential surface.}
\label{fig:K0FC}
\end{figure}

In the remaining part of this work, we will analyze the spectroscopic properties of the crystal which derive from this state dependent configuration change. We proceed by introducing normal modes that describe the vibrational motion of the ion crystal on the two potential surfaces which correspond to the electronically excited ion crystal and the crystal with all ions in the ground state (see sketch in Fig.~\ref{fig:scheme}c). This results in the vibrational Hamiltonian $H_\mathrm{vib}=\sum_i\hbar\omega_ia^{\dagger}_ia_i P_-+\sum_j\hbar\nu_jb^{\dagger}_jb_jP_+$. Here $a_i(b_j)$ are the normal mode annihilation operators on the lower (upper) potential surface with vibrational frequencies $\omega_i$ ($\nu_j$). $P_{\mp}=(1\mp\sigma_z)/2$ projects on the electronic ground state/Rydberg state of the central ion and involves the Pauli matrix $\sigma_z$. To simplify the discussion we assume that the central ion is already excited to the metastable $3D_{3/2}$ state, which can be done without affecting the crystal configuration. After including the Rydberg excitation laser on the $3D_{3/2}$-$nP_{1/2}$-transition with Rabi frequency $\Omega_{\rm{L}}$ the total Hamiltonian becomes $H=H_\mathrm{vib}+H_\mathrm{L}$ where
\begin{equation}
H_\mathrm{L}=-\frac{\hbar\delta}{2}\sigma_z+\frac{\hbar \Omega_{\rm{L}}}{2}\sum_{[n,m]}\left[A_{[n]}^{[m]}b^{\dagger}_{[m]}a_{[n]}\sigma^++\rm{h.c.}\right].
\label{eq:hamiltonian}
\end{equation}
and $\sigma_{\pm}=1/2(\sigma_x\pm i\sigma_y)$ (with $\left|3D_{3/2}\right>\equiv\left|\downarrow\right>$ and $\left|nP_{1/2}\right>\equiv\left|\uparrow\right>$). Here we have introduced the detuning $\hbar\delta=\hbar\omega_{\rm{L}}-(E_{nP_{1/2}}+\Delta E+E_{\rm{et}})$ where $\omega_{\rm{L}}$ is the laser frequency, $E_{nP_{1/2}}$ is the $nP_{1/2}$ state energy relative to the metastable $3D_{3/2}$ state,  $\Delta E$ is the energy shift of the two potential surfaces resulting from the structural change of the ions, and  $E_{\rm{et}}$ a constant energy shift due to the electron-trap coupling $H_{\rm{et}}$. The coefficients $A_{[n]}^{[m]}$ are the so-called Frank-Condon (FC) factors and given by the overlap integrals of the vibrational modes of the lower and excited potential surface. Here $[n]$ and $[m]$ are multi-indices containing all phonon occupation numbers of the lower/upper modes. They are also used for abbreviating $a_{[n]}=\prod_{n_i}a_{n_i}$, where $a_{n_i}$ is the annihilation operator corresponding to the $n$-phonon state of the $i$-th mode. The same notation applies to $b^{\dagger}_{[n]}$.

\begin{figure}
\centering
\includegraphics*[width=0.96\columnwidth]{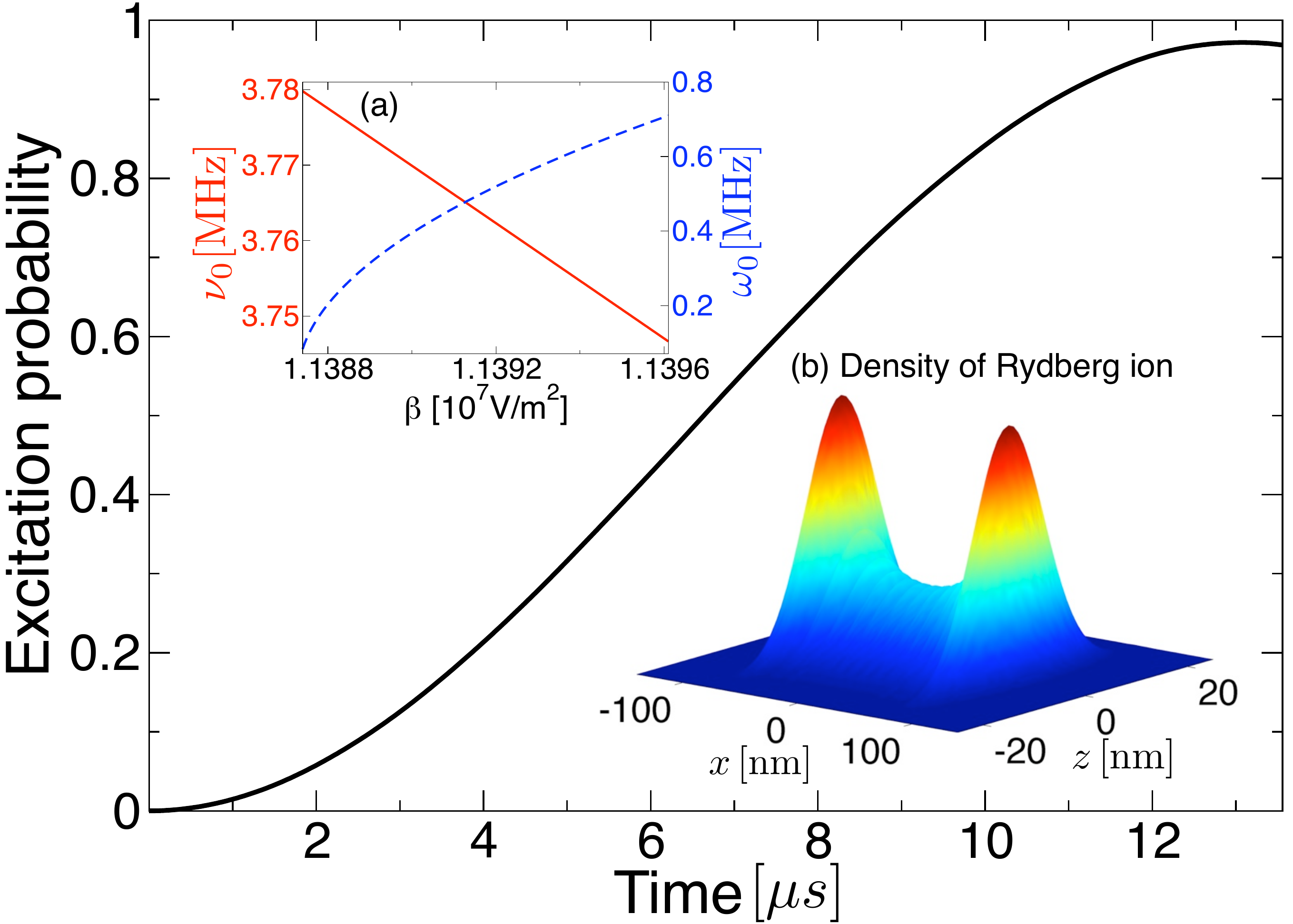}
\caption{Due to non-zero FC factors phonon states containing many vibrational quanta can be directly addressed. The data in the main figure shows the excitation probability of an $11$-phonon state in the upper potential surface. We have considered the $38P_{1/2}$ state of $^{40}$Ca$^+$, the laser is resonant with the desired phonon state and $\Omega_{\rm{L}}=2\pi\times 0.2$ MHz, $\beta=1.1392\times 10^7 \rm{V/m^2}$. The corresponding FC factor is $\approx 0.195$. The inset (a) shows the phonon energies of the lowest energy normal modes, $\omega_0$ (all ions in the electronic ground state) and $\nu_0$ (electronically excited crystal). The inset (b) shows the density profile (cut along $y=0$) of the Rydberg ion CM motion in the excited vibrational 11-phonon state. }
\label{fig:rydexcitation}
\end{figure}

The FC factors crucially depend on the mutual displacement $|\mathbf{K}_0|$ of the minima of potential surfaces. In the region (b) of the `phase diagram' in Fig.~\ref{fig:criticalbeta} where the configurations of the ground state and the excited crystal differ substantially, $|\mathbf{K}_0|$ can significantly exceed the harmonic oscillator length, typically $\sim10$ nm. This is illustrated in Fig.~\ref{fig:K0FC}a for the Rydberg $38P_{1/2}$ state. In Fig.~\ref{fig:K0FC}b we show the corresponding FC factors for transitions starting from the vibrational ground state of the lower potential surface, which are numerically calculated by a Gaussian integral method \cite{sharp64}.
We observe FC factors that indicate significant overlap with highly excited vibrational states of the upper potential surface. The calculation furthermore shows that only the lowest vibrational mode in each potential surface determines the FC factor. Higher modes are virtually identical and the FC factors for transitions changing their vibrational state are zero. The frequencies of the two lowest modes ($\omega_0$ and $\nu_0$) are shown in the inset (a) of Fig.~\ref{fig:rydexcitation}.

The FC factors in Fig.~\ref{fig:K0FC}b are a result of strong mechanical forces that accompany the Rydberg excitation of the ion crystal. These forces are substantial also for crystals with large vibrational frequencies $\sim$ MHz. This is a striking difference to the common situation encountered in ion traps where merely low-lying electronic states are laser excited. Here one is usually in the Lamb-Dicke regime and the direct excitation of high-lying vibrational states is strongly suppressed. We can use the emerging large mechanical forces to selectively populate high-lying vibrational state of the ion crystal. This is depicted in Fig.~\ref{fig:rydexcitation} for a state on the upper potential surface containing 11 vibrational quanta  which is addressed \emph{directly} from the vibrational ground state of the lower surface. The process is state-selective as long as the product of the FC factor and the excitation Rabi frequency $\Omega_{\rm{L}}$ is smaller than the phonon energy $\omega_0$. Here we chose a realistic Rabi frequency of $\Omega_{\rm{L}}=2\pi\times 0.2$ MHz \cite{schmidt11} which achieves a population of the 11-photon state with more than $97\%$ probability in $t\approx 13\, \mu s$. This process creates a `moving' Rydberg ion with a probability density shown in Fig.~\ref{fig:rydexcitation}b.

This situation is distinct from conventional cold neutral Rydberg systems which are often studied in the so-called `frozen' regime \cite{anderson98,mourachko98} and where motion eventually leads to decoherence caused by Penning ionization \cite{amthor07}. In our case the `vibronic dynamics' is coherent due to the absence of collisions and the long lifetime of $^{40}$Ca$^+$ ions in the $nP_{1/2}$-state. For $n>15$ this lifetime is $\tau_0 = 1.534\times(n-\delta_1)^{2.967}$ns, with the quantum defect $\delta_1\approx 1.436$. These numbers agree well with qualitative estimates found in Ref.~\cite{djerad91} and yield $\tau_0\approx 66\,\mu s$ for a $38P_{1/2}$ state.

In conclusion, the excitation of a cold ion crystal to high-lying electronic states was studied and shown to be accompanied by large coherent forces. This can find practical application in the realization of geometric quantum gates \cite{duan01} and the coherent separation of long ion chains. Excited cold ion crystals permit moreover the quantum simulation of coherent molecular phenomena and in principle also allow the inclusion of radiative transitions between `molecular' energy surfaces as tight traps can grant confinement even after the de-excitation of Rydberg states by incoherent photon emission.

We thank C. Ates and S. Genway for careful reading of the manuscript. Funding through EPSRC and by the EU (Marie Curie Fellowship) is gratefully acknowledged.

\end{document}